\documentclass[%
 aip,
 jmp,%
 amsmath,amssymb,
 reprint,%
]{revtex4-1}

\usepackage{graphicx}% Include figure files
\usepackage{dcolumn}% Align table columns on decimal point
\usepackage{bm}% bold math
\usepackage{color}
\usepackage{textcomp,gensymb} %for \degree command
\usepackage[colorinlistoftodos]{todonotes}

\usepackage{multirow}
\usepackage{siunitx}
\usepackage{float}

\AtBeginDocument{
\usepackage{booktabs}
}

\newcommand{\beginappendix}{%
        \setcounter{section}{0}
        \setcounter{table}{0}
        \renewcommand{\thetable}{S\arabic{table}}%
        \setcounter{figure}{0}
        \renewcommand{\thefigure}{S\arabic{figure}}%
     }

\begin{document}

\preprint{APS/123-QED}

\title[A Numerical Study of Three-Armed DNA Hydrogel Structures]{A Numerical Study of Three-Armed DNA Hydrogel Structures}

\author{Yair Augusto Guti\'errez Fosado}
\thanks{Both authors contributed equally to this manuscript.}
\affiliation{School of Physics and Astronomy, University of Edinburgh, Edinburgh EH9 3FD, Scotland, UK}

\author{Zhongyang Xing}
\thanks{Both authors contributed equally to this manuscript.}
\author{Erika Eiser}
\affiliation{Optoelectronics, Cavendish Laboratory, University of Cambridge, Cambridge CB3 0HE, UK}

\author{Magdalena Hudek}
\author{Oliver Henrich}
\email{oliver.henrich@strath.ac.uk}
\affiliation{Department of Physics, SUPA, University of Strathclyde, Glasgow G4 0NG, Scotland, UK}

\date{\today}% It is always \today, today,

\begin{abstract}
We present a numerical analysis of a DNA hydrogel that consists of three-valent Y-shaped DNA molecules. The building blocks self-assemble fully reversibly from complementary single-stranded DNA segments. We compare melting curves from both simulations with the oxDNA2 model and experiments and find excellent agreement. The morphology of the Y-DNA molecules is investigated when several alterations in the design are made. Adding inert nucleotides to the central core region of the Y-DNA molecules has a very minor effect on their overall geometry, whereas palindromic sequences in the sticky ends via which the individual Y-DNA molecules hybridize, have a profound influence on their relative twisting and bending angles. 
\end{abstract}

\maketitle

%%%%%%%%%%%%%%%%%%%%%%%%%%%%%%%%%%%%%%%%
%%%       Section 1: Introduction    %%%
%%%%%%%%%%%%%%%%%%%%%%%%%%%%%%%%%%%%%%%%
\section{\label{sec:introduction}Introduction}

% background of DNA self-assembly: features, applications

The astounding base-pairing mechanism in DNA is nature's robust way to encode the genetic information of all living organisms. At low temperature, two antiparallel biopolymer chains that consist of a complementary sequence of the four nucleic acids adenine, thymine, cytosine and guanine to assemble into duplexes. In eukaryotes the DNA duplexes are organized hierarchically and further compactified into chromatin fibers and chromosomes, whereas in prokaryotes DNA duplexes usually exist in from of ring-shaped plasmids.

DNA nanotechnology uses the hybridization of single-stranded DNA (ssDNA) into double-stranded DNA (dsDNA) duplexes to design self-assembling, reversible structures with specific target shapes and connectivity, so-called DNA motifs. These motifs emerge when the DNA backbone is reciprocally exchanged between different duplexes ~\cite{Seeman:2007}.
A well known example that uses this fundamental principle is the Holliday junction ~\cite{Holliday:1964}. It plays a key role in meiosis, genetic recombination and DNA repair mechanisms. In Holliday junctions one such exchange of backbones takes place. Motifs with more exchanges are also allowed, e.g. double and triple crossovers, or molecules with paranemic crossovers that exchange strands wherever the backbones of the duplexes are in close contact. This approach allows to design very complex three-dimensional structures, for instance DNA bricks that consist of ssDNA strands which are composed of four binding domains ~\cite{Ke:2012}. The bricks interact via simple local binding rules and are modular as they do not rely on a scaffold. This means that each brick can be added or removed independently.
Another striking example is a refinement of DNA origami that uses ssDNA and single-stranded RNA ~\cite{Han:2017}. The key challenge herein lies in finding a scheme that allows construction of complex, programmable single-stranded structures while maintaining a simple strand topology to avoid knotting, kinetic traps and ensure smooth folding. This was achieved by using partially complemented dsDNA and dsRNA.
A very different application that utilizes the base-pairing mechanism was realized in form of DNA-coated colloids, which are reversibly absorbed on oil droplets ~\cite{Joshi:2016}. Both the liquid-liquid interface and the colloidal surface are functionalized such that colloids can bind to the interface in a controlled manner through complementary DNA interactions.

DNA hydrogels are akin to Holliday junctions and have emerged as a distinct class of DNA material that is based on multi-valent structures with specific nucleotide sequence. DNA hydrogels attracted considerable attention due to their uses in controlled drug delivery, tissue engineering, for 3D cell cultures, cell transplant therapy and other biomedical applications. The development began with the first Y-shaped DNA that could be synthesized as it had sufficient purity and monodispersity to generate dendrimer-like DNA networks ~\cite{Li:2004}. The next experimental step consisted in replacing the ligase-based linking between the Y-DNA molecules with a base-pairing or hybridization mechanism ~\cite{Um:2006}.  Further improvements were made regarding the responsiveness to changes in the pH-value ~\cite{Cheng:2009} and temperature ~\cite{Xing:2011} when switching from the gel to the non-gel state. Particularly the latter improvement implies that the gel could rapidly form and dissolve without the requirement of any further chemical treatment. 

Since then new strategies have been suggested, e.g. to prepare supramolecular hydrogels with rationally designable and easily replaceable functionalization sites. These hydrogels are formed by DNA grafted polypeptide and X-shaped DNA linkers, cross-linked by precise DNA recognition. The entire gel formation process was accomplished within seconds under physiological condition ~\cite{Li:2015}. An intriguing development was achieved through the insertion of so-called i-motif sequences into the supramolecular DNA hydrogel structure ~\cite{Zhou:2016}. i-motifs are quadruplex structures different from the double helix,  which are formed by cytosine-rich DNA, similar to the G-quadruplex structures that guanine-rich DNA forms. Changes in the pH value trigger reversible conformational changes that modify the spatial distance between crosslinking points. This can be used to control the stiffness and elastic properties of the hydrogel ~\cite{Zhou:2016}.  It is worth mentioning that the microrheological properties and viscoelastic behavior of DNA hydrogels have only recently been probed extensively for the first time ~\cite{Stoev:2018, Fernandez-Castanon:2018,ZhongyangXing:2018}. We refer to ~\cite{Okay:2011} for a summary of different early synthetization strategies and their effect on the physical properties of DNA hydrogels. A more recent overview can be found ~\cite{Shao:2017}.

DNA self-assembly to produce nanoscopic particles with a controlled number of interacting terminations, providing the particles with valence ~\cite{Biffi:2013}. Experimental investigation of the collective behavior of DNA particles with specific valence shows that reducing the number of interacting sites results in a significant shrinkage of the gas–liquid coexistence region, with critical parameters decreasing as the valence is reduced and show an unconventional dynamic behavior in the proximity of the critical point.  

Such findings are relevant to answer fundamental physics issues and potentially to determine the stability region of new DNA-based materials and ask for numerical investigations. 
The first studies of DNA hydrogels and related materials started with tetra-valent DNA dendrimers and the formation of amorphous gel structures ~\cite{Starr:2006}. This early bottom-up approach considers the individual components, which are designed to assume particular tertiary structures with the aim of self-assembly into quaternary structures.
Owing to much refined computational models, tetra-valent DNA nanostars were found to form a thermodynamically stable equilibrium gel ~\cite{Rovigatti:2014a, Rovigatti:2014b, Locatelli:2017}. This is in stark contrast to atomic and molecular network formers, in which the disordered liquid is always metastable with respect to some crystalline phase. This unconventional behavior arises from the large arm flexibility of the DNA nanostars, a property that can be tuned by design. 
Hence, the appropriate selection and use of DNA–DNA interactions makes it possible to generate bulk quantities of colloidal nanoparticles that can be used for designing and realizing self-assembled soft materials with unconventional properties.  ~\cite{Bomboi:2016}
In fact, the situation is more complex. Recently, it has been demonstrated that by relaxing the fixed-valency constraint, flexible DNA junctions can indeed crystallise \cite{Brady:2017, Brady:2018} and that seemingly minor changes in nano- structure or buffer conditions lead to substantial differences in the structure of the network phases, giving evidence that flexibility can not only be an acceptable characteristic, but also an essential feature for successful crystallisation of amphiphilic DNA motifs \cite{Brady:2019}.

In this work, we build on the proven success of the oxDNA2 model to describe multi-valent DNA structures and study the melting behavior and morphology of the Y-DNA building blocks of a three-valent DNA hydrogel system where the individual molecules hybridize via sticky ends to form the percolating network structure. 
Our aim is to gain a deeper understanding, how the sequence may influence local structural features that can have a determining effect on larger length scales. These results will form a stepping stone for large-scale simulations of many such Y-shaped building blocks.
We draw on the oxDNA2 model for coarse-grained simulation of DNA ~\cite{Ouldridge:2011, Sulc:2012, Snodin:2015}, which has been recently ported into the LAMMPS code (Large Scale Molecular Massively Parallel Simulator) ~\cite{LAMMPS,LAMMPSWebsite,oxDNALAMMPS} and is now amongst other LAMMPS-based CG DNA models ~\cite{Hinckley:2013, Fosado:2016}.
The paper is organized as follows: In section \ref{sec:mm}Section 2 we introduce the computational and experimental methods and provide details of the data analysis. In section \ref{sec:results} we compare the melting curves of Y-DNA molecules in simulation and experiments and find both in excellent agreement. We characterize the morphology of the molecules study its dependence on different design choices. While introducing inert bases into the central core region of the Y-DNA molecule has only a vary minor effect on their shape, modifications in the sticky ends seem to be crucial to control the bending and twist angle between neighboring building blocks. In section \ref{sec:conclusion} we summarize our findings.

%%%%%%%%%%%%%%%%%%%%%%%%%%%%%%%%%%%%%%%%
%%%   Section 2: Models and Methods  %%%
%%%%%%%%%%%%%%%%%%%%%%%%%%%%%%%%%%%%%%%%
\section{\label{sec:mm}Models and Methods}

%%%%%%%%%%%%%%%%%%%%%%%%%%%%%%%%%%%%%%%%
\subsection{Simulation Methods}
%%%%%%%%%%%%%%%%%%%%%%%%%%%%%%%%%%%%%%%%

The oxDNA2 model ~\cite{Ouldridge:2011} is a coarse-grained model based on experimental data, in which each nucleotide is represented by a rigid body with distinct interaction sites - the interactions between the interaction sites are pairwise additive. These interactions account for: the excluded-volume interactions between nucleotides, the connectivity of the phosphate-sugar backbone and the stacking, cross-stacking and coaxial stacking forces between nucleotides and, finally, for the hydrogen bonding between complementary base pairs (bp).

The simplest interaction is the backbone connectivity, which is modeled with FENE (finitely extensible nonlinear elastic) springs acting between the backbone interaction sites. The excluded volume interaction is modeled with truncated, shifted and force-smoothed Lennard-Jones potentials between backbone sites, base sites and between the backbone and base sites. The (parabolic) smoothing ensures that the force goes to zero continuously at the cut-off distance. The potential parameters that we use are given in~\cite{Tom-thesis, Snodin:2015}. The hydrogen bonding interaction consists of smoothed, truncated and modulated Morse potentials between the hydrogen bonding site.
The stacking interaction falls into three individual sub-interactions: the stacking interaction between consecutive nucleotides on the same strand as well as cross-stacking and coaxial stacking between any nucleotide in the appropriate relative position.
It is worth emphasizing that the duplex structure is not specified or imposed in any other way, but emerges naturally through this choice of interactions and their parameterisation. This is another strength of the oxDNA model and permits an accurate description of both ssDNA and dsDNA.
The stacking interactions are modeled with a combination of smoothed, truncated and modulated Morse, harmonic angle and harmonic distance potentials.
All interactions have been parameterized to match key thermodynamic properties of ssDNA and dsDNA such as the longitudinal and torsional persistence
length or the melting temperature of the duplex ~\cite{Holbrook:1999,SantaLucia:2004}.

The Langevin Dynamics simulations that we report here were performed with the oxDNA2 model ~\cite{Snodin:2015}, which is the latest and improved version in the series of oxDNA models. It features not only the implicit ions via a Debye-H\"uckel electrostatic interaction, but reproduces also the correct structure of dsDNA with major and minor groove. This is achieved through a modification of the relative position of the backbone and stacking/hydrogen bonding interaction sites. For simplicity, we will simply refer to the model as oxDNA.

%%%%%%%%%%%%%%%%%%%%%%%%%%%%%%%%%%%%%%%%
\subsection{DNA Hydrogel Molecules}
%%%%%%%%%%%%%%%%%%%%%%%%%%%%%%%%%%%%%%%%

The building blocks of the DNA hydrogel consist of Y-DNA  molecules that are taken from the system in Ref.~\cite{ZhongyangXing:2018}. This binary system has only two types of Y-shaped DNA nanostars, whose sticky ends are complementary so that different types can bind together via DNA hybridization. We refer to these two Y-shaped DNA nanostars as $S$-DNA and $S'$-DNA molecules, respectively. Each is constructed from three partially complementary single-stranded (ss) oligonucleotides (named as $S_1$, $S_2$ and $S_3$ for T-DNA and $S'_1$, $S'_2$ and $S'_3$ for $S'$-DNA molecules). A single oligonucleotide is 46 bases long, with three main functional sections: (i) the main core (30 bases long) that forms the dsDNA arms (15-bases for each arm); (ii) the sticky ends (12 bases long) that is used for crosslinking other Y-shapes; (iii) the free joint (4 bases long) that bridges the main core and the sticky end, providing flexibility between two building blocks in conjunction.  The double-stranded arms of all nanostars are formed by the same sequences. The details of the sequences are summarized in Table ~\ref{table:sequence}. 

\begin{table}[h!]
\centering
\resizebox{\columnwidth}{!}
{
\begin{tabular}{*5c}
\toprule
ssDNA    & Sticky end       & Free joint & \multicolumn{2}{c}{Core} \\
\midrule
 {}      &              {}                            &              {}           &         Segment I   &         Segment II           \\
 %PART OF THE TABLE FOR THE T-MOLECULE
 \begin{tabular}{@{\ }l@{}}
 $S_{1}$ \\ $S_{2}$ \\ $S_{3}$
 \end{tabular}

 & $5^{\prime}-$ TGTCACTCACAG & TTTT & 

 $\left\{\begin{tabular}{@{\ }l@{}}
 TGGATCCGCATGATC \\ TACTTACGGCGAATG \\ AGGCTGATTCGGTGT  \end{tabular}\right.$ & 

 \begin{tabular}{@{\ }l@{}}
    CATTCGCCGTAAGTA $-3^{\prime}$ \\ ACACCGAATCAGCCT $-3^{\prime}$ \\ GATCATGCGGATCCA $-3^{\prime}$  \end{tabular}\\
     &                                            &                           &                     &             \\
 %PART OF THE TABLE FOR THE T'-MOLECULE
 \begin{tabular}{@{\ }l@{}}
 $S_{1}^{\prime}$ \\ $S_{2}^{\prime}$ \\ $S_{3}^{\prime}$
 \end{tabular}

 & $5^{\prime}-$ CTGTGAGTGACA & TTTT & 

 $\left\{\begin{tabular}{@{\ }l@{}}
 TGGATCCGCATGATC \\ TACTTACGGCGAATG \\ AGGCTGATTCGGTGT  \end{tabular}\right.$ & 

 \begin{tabular}{@{\ }l@{}}
    CATTCGCCGTAAGTA $-3^{\prime}$ \\ ACACCGAATCAGCCT $-3^{\prime}$ \\ GATCATGCGGATCCA $-3^{\prime}$  \end{tabular}\\

\bottomrule
\end{tabular}
}
\caption{The bases of the three oligonucleotides that form the $S$ and $S'$ molecule, respectively. The segments I and II of each ssDNA molecule are designed to hybridize and form the Y-shape core of the molecule. For instance, segment I on $S_1$ is complementary to segment II on $S_3$. Correspondingly, the T' molecule is made by $S'_1$, $S'_2$ and $S'_3$. $S$ and $S'$ molecules have the same core sequences, but their sticky ends are palindromic and complementary.}
\label{table:sequence}
\end{table}

\begin{figure}[htpb]
\centering
\includegraphics[width=0.5\textwidth]{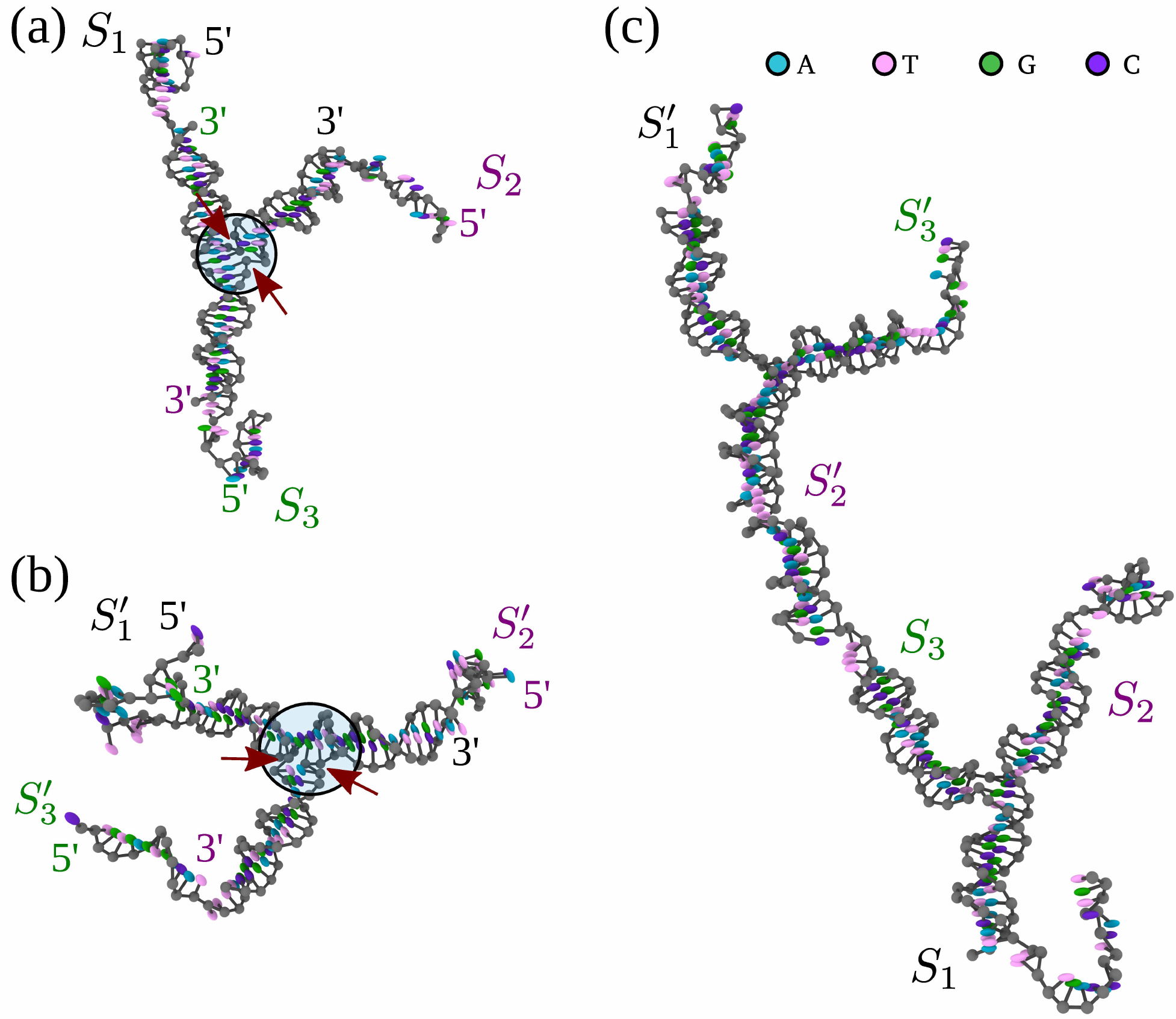}
\caption{(\textbf{a}) Snapshot from the simulations of an equilibrated $S$ composite molecule. The three ssDNA molecules $S_{1}$, $S_{2}$, $S_{3}$ and their directionality (from $5'$ to $3'$) are shown with the same color. (\textbf{b}) Snapshot of an equilibrated $S'$ molecule. 
The arrows in the insets of a and b mark regions where we expect non-canonical base pairing. (\textbf{c}) The sticky ends of the $S$ and $S'$ molecules hybridize and form the building blocks of a large-scale network. The color code in all images depicts the sugar-phosphate backbone sites as gray beads whereas the bases are represented by an ellipsoid according to a color scheme with A (blue), T (pink), G (green) and C (purple).}
\label{fig:panel1}
\end{figure}

Fig.\ref{fig:panel1}(a)-(b) shows snapshots from the simulations of the $S$ and $S'$ molecules at low temperature where all nucleotides in the core sections are hybridized. A duplex formed by a $S$ and $S'$ molecule can be seen in Fig.\ref{fig:panel1}c). The oxDNA model allows only for canonical base pairs between A-T pairs and C-G pairs (note that oxDNA allows only for Watson-Crick base pairs and Hoogsteen base pairs are not supported). This in combination with the right-handed double helix leads to tightly twisted oligonucleotides at the center of the $S$ or $S'$ molecules. A more detailed model as it would be e.g. used in atomistic simulations incorporates as well the possibility of forming non-canonical base pairs, which are either weaker polar hydrogen bonds, interactions between groups of atoms or hydrogen bonds between pairs of nucleotides other than the above mentioned ones. In oxDNA, this absence of non-canonical base pairs leads to conformations that can feature a kink at the center of the compound molecule where the three arms meet. However, it should be noted that kinks have also been observed in direct comparisons of oxDNA with atomistic simulations of DNA minicircles ~\cite{Sutthibutpong:2016}. Direct comparisons of DNA nanostar with oxDNA conformations that also featured kinks and less detailed cryoTEM imagery showed excellent overall agreement ~\cite{Schreck:2016}. Hence, the kinks we observe may not be classified as a pure artefact of the coarse-grained representation.

To release some of this frustration we studied as well modified molecules that contained additional nucleotides at the central core. These additional nucleotides were chosen to be inert, i.e. they do not participate in any hydrogen bond. We refer to these modified molecules as $S \, n_1 S_1 \, n_2 S_2 \, n_3 S3$ where $n_1, n_2, n_3$ is the number of additional inert nucleotides in the corresponding arm. The equivalent notation with primes applies to $S'$-DNA molecule. 
 
%%%%%%%%%%%%%%%%%%%%%%%%%%%%%%%%%%%%%%%%
\subsection{Data Analysis \label{sec:dataanalysis}}
%%%%%%%%%%%%%%%%%%%%%%%%%%%%%%%%%%%%%%%%

We take the configurations of the fully assembled Y-DNA molecule at equilibrium for further analysis to study the geometric features of the system. The emphasis is on the angle profile between two neighboring dsDNA arms and the planarity of a single Y-shaped molecule. 

\begin{figure}[htpb]
\centering
\includegraphics[width=0.5\textwidth]{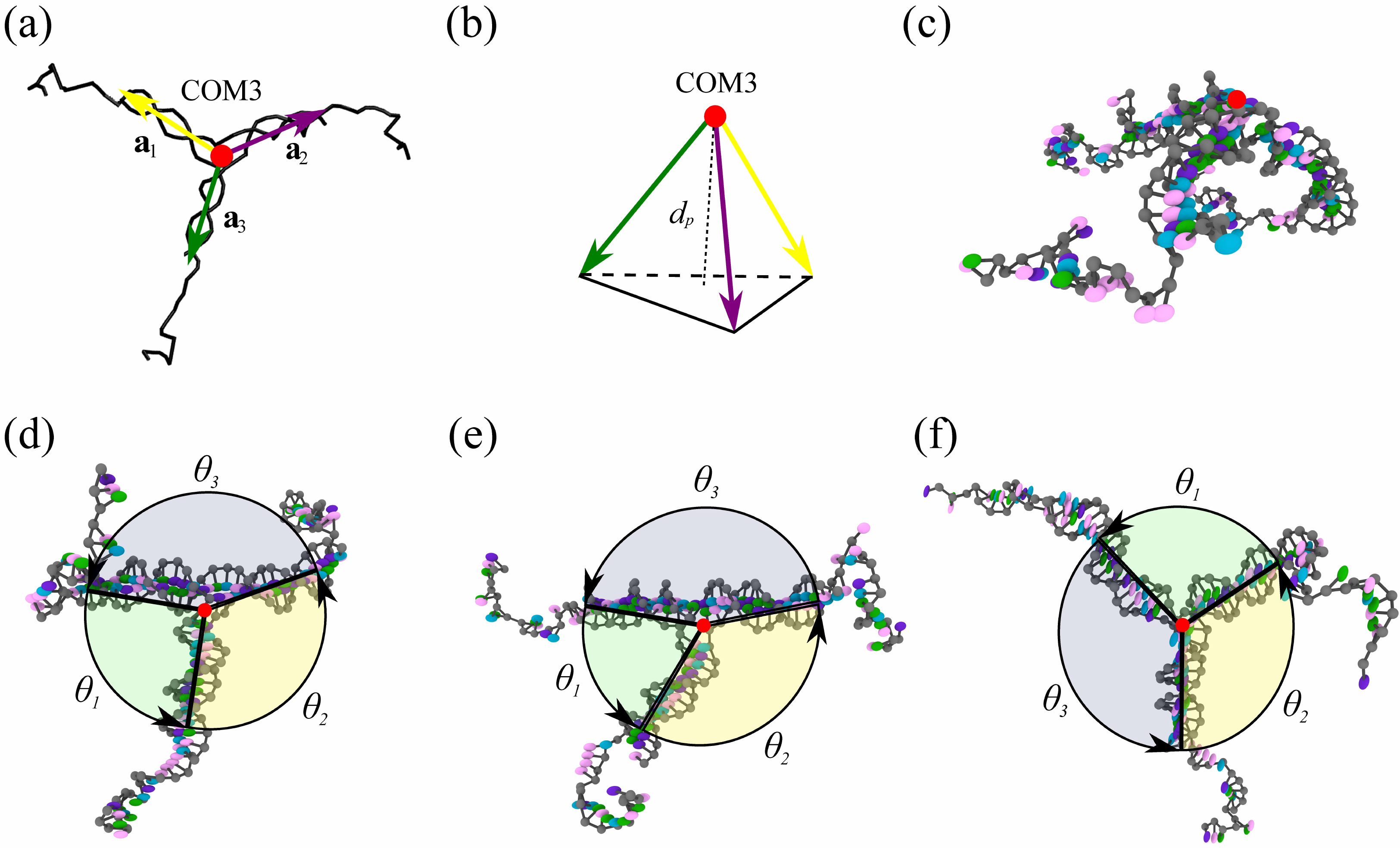}
\caption{(\textbf{a}) Schematic representation of the direction $\mathbf{a}_{i}$ of the three dsDNA arms in a Y-DNA molecule. (\textbf{b}) The planarity of the molecule is related to the distance $d_{p}$ between the plane touching the ends of the arms and COM3. (\textbf{c}) Example of a non-planar molecule with $d_{p}=0.5$. (\textbf{d})-(\textbf{f}) show planar molecules of Type-I, Type-II and Type-III (T-, T-like- and Y-shape), respectively. Then angles between two different vectors $\mathbf{a}_{i}$:  $\theta_{1}$, $\theta_{2}$ and $\theta_{3}$ are also shown. They are labeled according to their magnitude, with $\theta_{3}>\theta_{2}>\theta_{1}$.}
\label{fig:panel2}
\end{figure}

To acquire the angle profile, it is necessary to first define the vector that represent each dsDNA arm (called `arm vector' in the later context). Ideally, the center of mass (COM) of the Y-DNA molecule would serve as starting point of each arm vector, and the center of the last base-pair on each arm as the corresponding end point (Fig.~\ref{fig:panel2}(a)). However, due to the massive fluctuations of the ssDNA overhangs and the occasional temporal denaturation of nucleotides in the core part of the Y-DNA molecule, it is actually very difficult to extract an unambiguous COM as it always deviates from the geometric center of the core. For the sake of simplicity and accuracy, we therefore re-define the starting point of the arm vector and choose instead the COM of the most inner non-denatured base-pairs on each arm. We refer to this definition of the COM as `COM3'. Note that there is always an offset between the COM of the Y-shape and COM3 at any given time, unless the Y-shaped structure is fully hybridized and the arms and overhangs are nicely stretched. Fig.~\ref{fig:panel2}(a) shows the schematic of the three arm vectors ($\mathbf{a}_{1}$, $\mathbf{a}_{2}$ and $\mathbf{a}_{3}$) in one Y-shaped DNA.  

In order to characterize the planarity of the individual Y-DNA molecule we calculate the normalized distance $d_p$ from COM3 to the plane that is defined by the end points of the vectors $\mathbf{a}_{1}$, $\mathbf{a}_{2}$ and $\mathbf{a}_{3}$ (see Fig.~\ref{fig:panel2}(b)).
Obviously, $d_{p} \in [0,1]$, whereas the molecule is perfectly planar when $d_p = 0$. Larger values of $d_{p}$ mean the planarity is less pronounced,  but some aspects are still retained within a certain range. In the following text, we will refer to a Y-shape structure whose $d_{p}\leq 0.175$ as a \textit{planar} system and otherwise a \textit{non-planar} system. Fig.~\ref{fig:panel2}(c) shows for instance a snapshot of a non-planar configuration where $d_{p}=0.5$. 

We classify the molecules in three categories by the relative orientation of their dsDNA arms and magnitude of the angles $\theta_{k}$ ($k$ = 1, 2, 3) between them. $\theta_{i}$ is calculated by using the dot product 
$\theta_k = \cos{(\mathbf{a}_{i} \cdot \mathbf{a}_{j})}, ~~i \neq j \in \{1,2,3\}$. 
We sort the set $\theta_{k}$ in ascending order so that $\theta_{1} < \theta_{2} < \theta_{3}$. 
Fig.~\ref{fig:panel2}(d) - (f) illustrate three typical shapes that were found during the analysis. 
In Fig.~\ref{fig:panel2}(d) $\theta_{3} \approx$ \ang{180} whereas $\theta_{1} \approx \theta_{2} \approx$ \ang{90}. We refer to this conformation as \textit{Type-I} or \textit{T-shaped}. 
We consider also molecules whose angles fluctuate like $\theta_{3}\in [\ang{160}, \ang{220}]$ and $\theta_{2}, theta_{1} \in [\ang{60},\ang{120}]$ to be of that type. 
Fig.~\ref{fig:panel2}(e) shows a molecule with $\theta_{3} \approx$ \ang{180}, but $\theta_{2}>\ang{120}$ and $\theta_{1}<\ang{60}$, a shape we label as \textit{Type-II} or \textit{T-like-shaped}. Finally, the most common conformation is characterized by $\theta_{3}<\ang{160}$ as is shown in Fig.\ref{fig:panel2}(f). This we call \textit{Type-III} or \textit{Y-shaped}.

%%%%%%%%%%%%%%%%%%%%%%%%%%%%%%%%%%%%%%%%%%%
%%% Section 3: Results & Discussions    %%%
%%%%%%%%%%%%%%%%%%%%%%%%%%%%%%%%%%%%%%%%%%%
\section{\label{sec:results}Results and Discussions}

\subsection{Melting curves of hybridized three-valent DNA complex}

We studied the melting behavior of the $S$ and $S'$ composite molecules by determining the hybridization yield, i.e. the number of nucleotides that form base pairs. While the experiments relate the absorbance of light at a wavelength of $260$ nm to the level of hybridization, the fraction $\phi$ of denatured base pairs can be directly accessed in the simulations. We started from a completely hybridized and equilibrated molecule (either $S$ or $S'$), performed a sudden quench to the target temperature and kept track of the fraction of denatured base pairs until a new equilibrium state had been reached, which we determined through a new steady value of $\phi$. 

\begin{figure*}[htpb]
\centering
\includegraphics[width=1.0\textwidth]{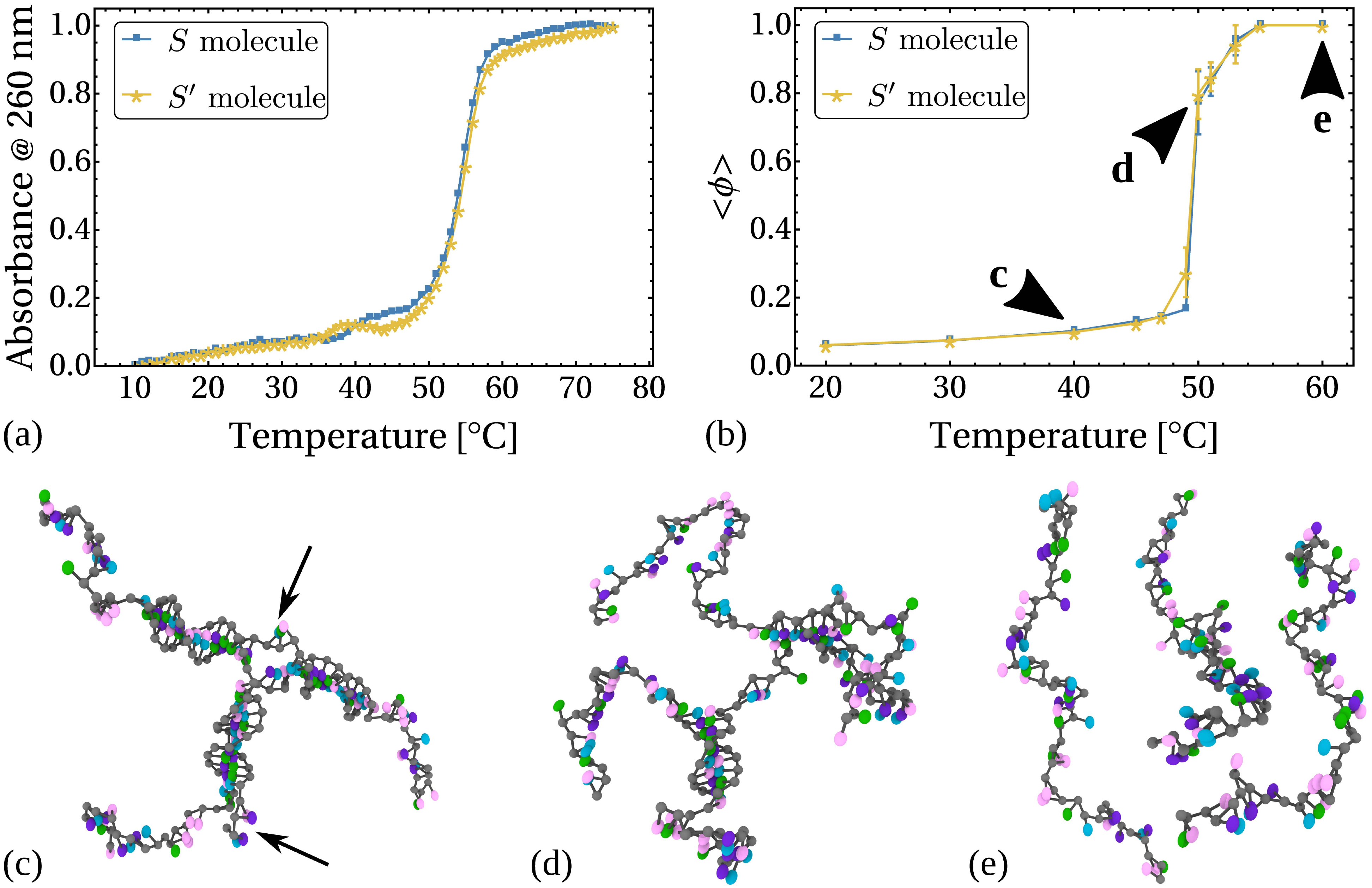}
\caption{(\textbf{a}) Experimental melting curves (absorbance at $260$ nm as a function of temperature) for $S$ and $S'$ molecules. (\textbf{b}) The melting curves (fraction of denatured base pairs $\phi$ as a function of temperature) obtained from simulation. Averages over 10 replicas and $10^{5} \tau_{LJ}$ were taken. (\textbf{c})-(\textbf{e}) Snapshots of typical configurations at $40\degree$C, $50\degree$C and $60\degree$C, respectively. The arrows in (c) indicate the onset of denaturation in regions at the core of the composite molecule and at end of one arm.}
\label{fig:panel3}
\end{figure*}

Profiles of the melting curves at a NaCl concentration of 200 mM were obtained from experiments and simulations. Both show very good qualitative agreement and are displayed in Figs. \ref{fig:panel3}(a) and \ref{fig:panel3}(b), respectively. 
In the lower temperature range the simulation curves are flattened out while the experimental curves are still decaying. This might be due to an electrostatic effect of the ssDNA overhangs that do not contribute to form the dsDNA cores, which shifts the hybridization free energy. This was demonstrated to affect the melting transition ~\cite{DiMichele:2014}, but is currently not included in oxDNA2.

The critical melting temperature at which half of the base pairs are melted is about $53\degree$C in experiments and $49.5\degree$C in simulations, a deviation of 1.3\% in absolute temperature. The denaturation trends of both $S$ and $S'$ molecules is very similar, indicating that the effect of the sticky ends -- the sole difference between the two types of molecule -- is almost negligible.

A longstanding observation from experiments is that the denaturation process in linear dsDNA nucleates from the ends of the strand ~\cite{Beers:1967}. It is interesting to see that -- contrary to linear dsDNA -- our $S$ and $S'$ Y-DNA molecules do not always begin to denature from the end of their arms, but sometimes straight from the central part of the molecule (see Fig.~\ref{fig:panel3}(c)). We attribute this to the relatively large frustration in central region of the molecule where elastic forces compete with the base pairing and weaken the hybridization mechanism. Hence, the question arises to what extent the structural properties of the molecule can be influenced through a modified design of the central region. 

\subsection{Morphology}

We studied the temperature dependence of the unmodified $S$ molecule.
Since the core-structures of $S$ and $S'$ molecules are the same, all reported observations can be extended to $S'$ molecules or its variants. 
We looked only at temperatures below the melting temperature around $49\degree$C to ensure the core structure is not fully denatured as the arm vectors could not be defined at higher temperatures. The results are summarized in Table~\ref{table:times}. 

At all temperatures between $20\degree$ C and the melting temperature around $49\degree$C where the molecules denature, the $S$ molecule spends most of its time in a planar configuration. The dominating conformation is the Type-III or Y-shape (between 53 and 55\% of the time).
As the temperature increases, thermal fluctuations of the arms increase as well and result on average in a less planar configuration.
 It is, however, interesting to observe that these fluctuations do not significantly affect the fraction of time that the molecule is in a Type-III conformation, but that the planarity decreases mainly because the molecules spend less time in a Type-I conformation. 

For instance at $20\degree$C the angles between three arms are on average $\langle \theta_{1}\rangle=90\degree$, $\langle\theta_{2}\rangle=118.5\degree$ and $\langle \theta_{3}\rangle=147.5\degree$, respectively. Note that because the molecule is not exactly planar ($d_{p}=0$) the three angles do not add up to $360\degree$. 
When the temperature increases to $49\degree$C, these averages become $\langle\theta_{1}\rangle=92$, $\langle\theta_{2}\rangle=120$ and $\langle\theta_{3}\rangle=145$.
So $\theta_1$ and $\theta_2$ increase, while $\theta_3$ slightly decreases. The $S$-molecule, however, remains still during 53\% of the time in the Type-III or Y-shaped configuration. 

\begin{table}[htpb]
\centering
\begin{tabular}{*5c}
\toprule
Temperature [\degree C] & \multicolumn{4}{c}{Time spent in a configuration[$\%$]}\\
\midrule
     &  Planar  &  Type-I &  Type-II & Type-III\\
20   &  83      &  25 &  3 & 55 \\
30   &  80      &  22 &  4 & 54 \\
40   &  77      &  20 &  4 & 53 \\
45   &  73      &  17 &  3 & 53 \\
49   &  70      &  14 &  3 & 53 \\
\bottomrule
\end{tabular}
\caption{Percentage of time that a $S$ molecules exist in planar, T-shaped (Type-I),  T-like-shaped (Type-II) or Y-shaped (Type-III) configuration (see also section \ref{sec:dataanalysis}). The total time that a molecules can be classified as planar (i.e. $d_{p} \leq 0.175$) is the sum of the three subcategories. The simulation data was obtained by averaging over time and ten independent configurations at each temperature setting. Note that for temperatures above $49\degree$C the duplexes that constitute the arms of the composite molecule denature and are not anymore well defined. Virtually similar data has been obtained for the $S'$ molecules.}
\label{table:times}
\end{table}

\subsection{Effect of introducing inert bases}\label{inert_bases}

\begin{figure*}[htpb]
\centering
\includegraphics[width=1.0\textwidth]{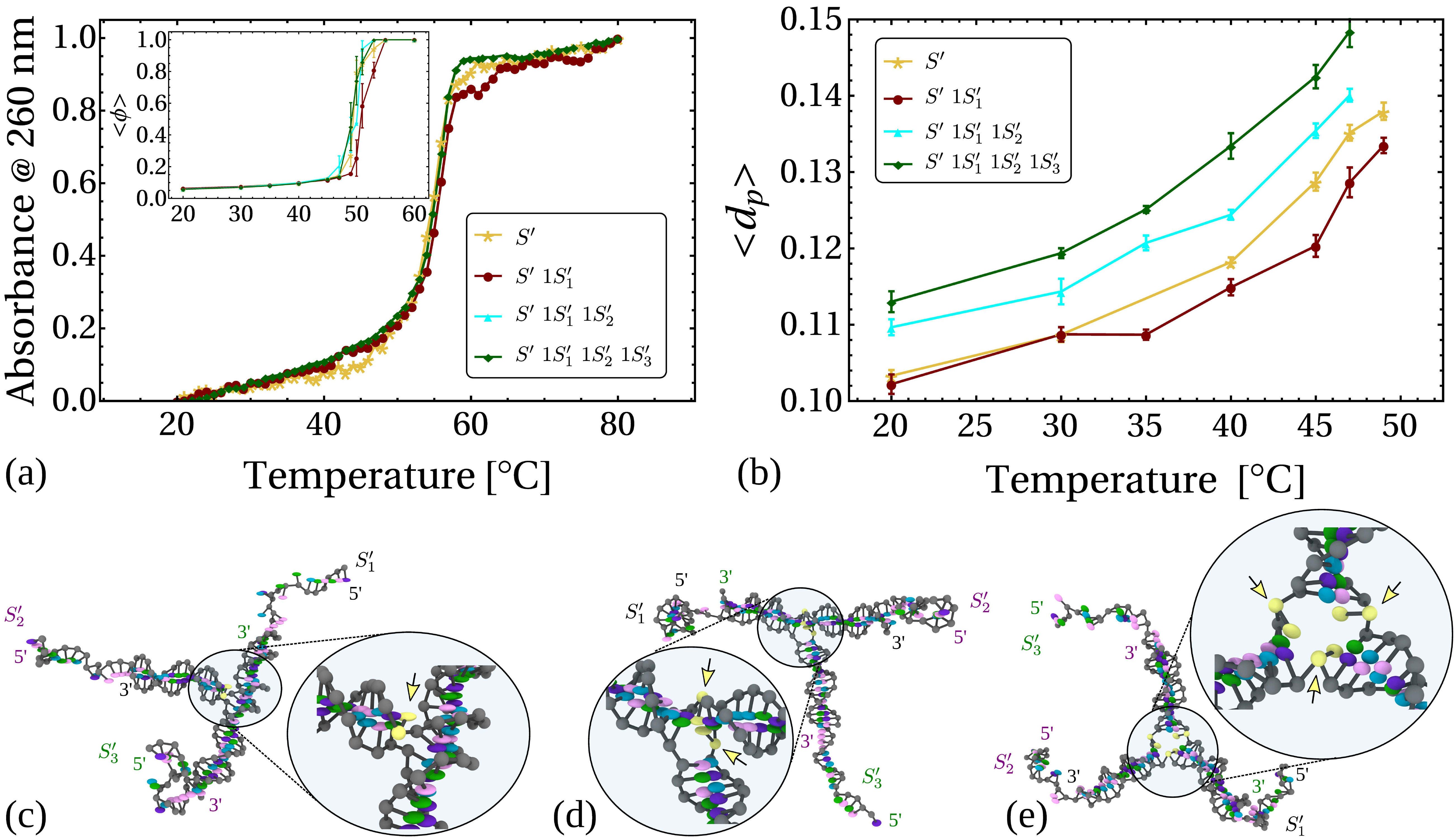}
\caption{(\textbf{a}) Effect of inert bases on the melting curves observed in experiments and simulations (inset). The label indicates the number of extra nucleotides added and in which of the three ssDNA oligonucleotides were placed. For example, $S'$ 1$S'_{1}$  (shown in red) represents the case where one extra nucleotide was added between the two segments I and II of the  $S'_{1}$ oligonucleotide. (\textbf{b}) The average distance $\langle d_{p}\rangle$ as function of the temperature of the system. (\textbf{c})-(\textbf{e}) Snapshots with nucleotides inserted into one (1$S'_1$), two (1$S'_1$  1$S'_2$) or all three oligonucleotides (1$S'_1$ 1$S'_2$ 1$S'_3$).}
\label{fig:panel4}
\end{figure*}

We investigated how the melting temperature and morphology changes when additional inert nucleotides (i.e. nucleotides that cannot form base pairs) are introduced into the central core region of the $S'$ molecule. 
Fig.\ref{fig:panel4}(a) shows the melting curves of the unmodified $S'$ molecule as well as those of modified molecules that contain different numbers of added inert nucleotides between segments I and II of the oligonucleotides (see Tab.\ref{table:sequence}).
The melting curves for one ($S'$ 1$S'_1$), two ($S'$ 1$S'_{1}$ 1$S'_{2}$) and three ($S'$ 1$S'_{1}$ 1$S'_{2}$ 1$S'_{3}$) additional bases feature are slightly different from that of the original $S'$ molecule. There is, however, a subtle, non-monotonous behavior as the melting temperature of $S'$ 1$S'_1$ is overall the highest, although all are more or less within the assumed range of uncertainty. This shows that there is no significant difference in DNA thermodynamics due to the introduction of inert bases in the core region. The experimentally acquired melting transition occurs between $54$ $\sim$ $56\degree$C, which is quantitative close to the simulation prediction between $49.5\degree$C and $51\degree$C. 
Fig.\ref{fig:panel4}(b) depicts the average distance $\langle d_p\rangle$ between COM3 and the end of the arms. There is a general trend that the composite molecules become less planar with increasing temperature, regardless of the number of extra nucleotides. The molecule with one additional nucleotide turns out to have also the lowest value of $\langle d_{p}\rangle$ and is therefore the most planar one.
Snapshots from the simulation are shown in Fig.\ref{fig:panel4} (c)-(e) for an increasing number of inserted nucleotides. This strictly non-monotonous trend, which is also seen in the simulation value of the melting temperature, could be explained in the following way: The larger contour length of the $S'_1$ oligonucleotide with one additional base opens up the central region of the molecule and reduces some of the structural frustration that manifests itself in form of large degrees of twist and/or kinks, as shown in Fig.~\ref{fig:panel1}(a)-(b). When inserting more than one nucleotide the configuration builds up stress again when forming base pairs, which leads to higher melting temperatures (Fig. \ref{fig:panel4}(a)) and less planar molecules (Fig. \ref{fig:panel4}(b)).

From Tab.\ref{table:timesinert} we see that adding inert nucleotides has in general only a minor effect on the planarity and shape of the molecules with the Type-III or Y-shaped configuration still being the most common one. The average angles of the original Y-shaped $S'$ molecule are $\langle\theta_{1}\rangle=90.5\degree$, $\langle\theta_{2}\rangle=118.2\degree$ and $\langle\theta_{3}\rangle=147.5\degree$ and change only to $\langle\theta_{1}\rangle=90.2\degree$, $\langle\theta_{2}\rangle=120\degree$ and $\langle\theta_{3}\rangle=145.5\degree$ when an extra nucleotides is inserted into each of the arms. 

\begin{table}[htbp]
\centering
\begin{tabular}{l*4c}
\toprule
Molecule                & \multicolumn{4}{c}{Time spent in a configuration[$\%$]}\\
\midrule
                        {}         &  Planar  & T-shape  &  T-like-shape & Y-shape\\
                        $S'$       &  83      & 25 	 &  3           & 55 \\
$S'$ 1$S'_{1}$                     &  83      & 23 	 &  6           & 54 \\
$S'$ 1$S'_{1}$ 1$S'_{2}$           &  80      & 20 	 &  4           & 56 \\
$S'$ 1$S'_{1}$ 1$S'_{2}$ 1$S'_{3}$ &  79      & 17       &  5           & 57 \\
\bottomrule
\end{tabular}
\caption{Percentage of time that $S'$ molecules with various numbers of additionally inert nucleotides exist in a particular Type-I,  Type-II or Type-III configuration (c.f. Section \ref{sec:dataanalysis}). The data was obtained by averaging over time and ten independent configurations at a temperature of $20\degree$C. The notation in the molecule column represent the number of additional T nucleotides and the specific oligonucleotide into which they have been inserted. For instance $S'$ 1$S'_1$ indicates that one additional nucleotide has been inserted between the two segments of the oligonucleotide $S'_1$.}
\label{table:timesinert}
\end{table}

\subsection{Effect of flexible joints}
\label{sec:effectFJ}

\begin{figure*}[htpb]
\centering
\includegraphics[width=1.0\textwidth]{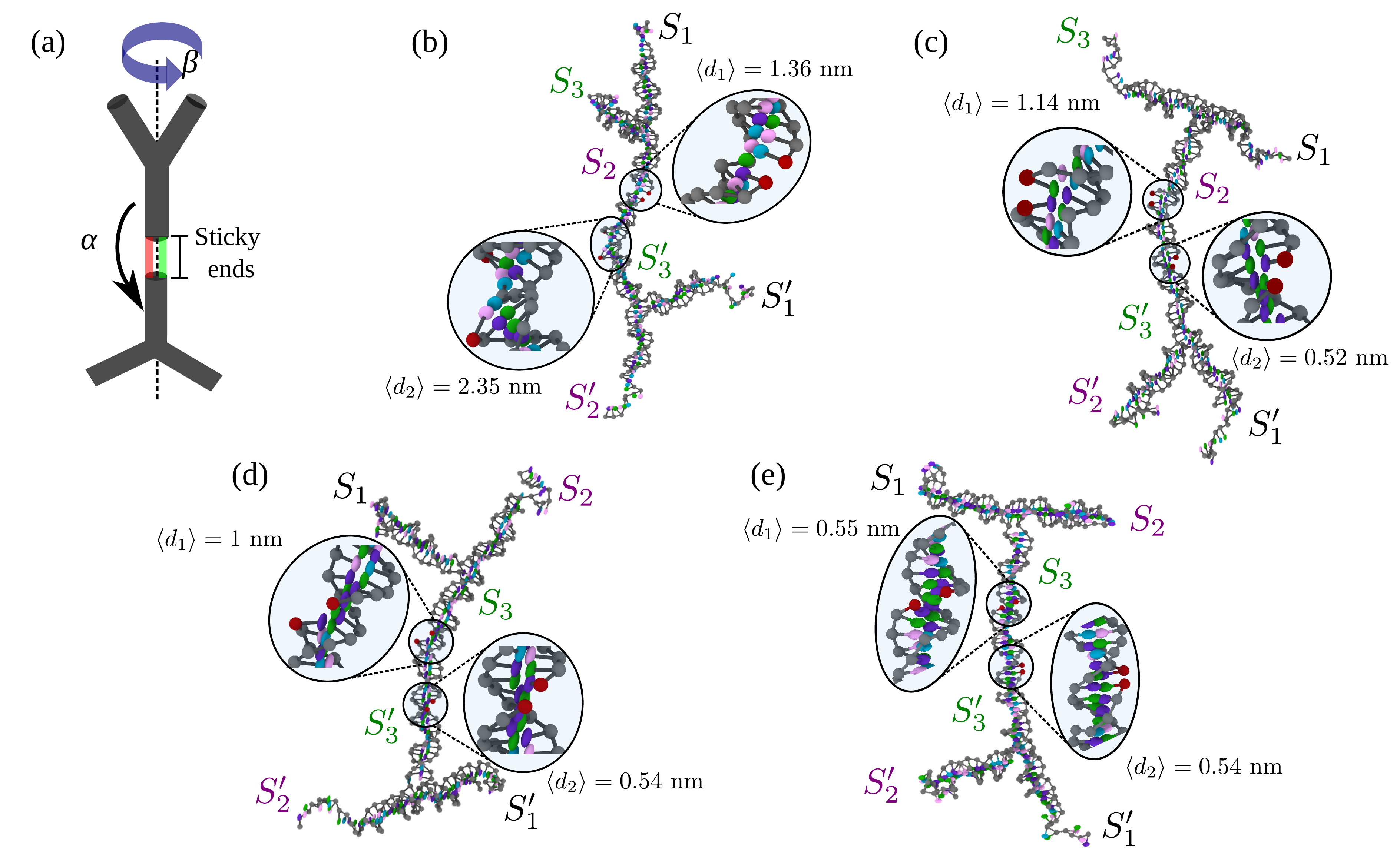}
\caption{(\textbf{a}) Sketch of the relative orientation of two complementary $S$ and $S'$ molecules, determined by the bending angle $\alpha$ between the two linked arms and the twist angle $\beta$ between the two dihedral planes of the molecules. (\textbf{b}) 2D-sketch and simulation of the two molecules without flexible joint. At the right a snapshot from the simulations is shown where the two nicked sections can exist either in the open state or in the closed state. Panels (\textbf{c})-(\textbf{e}) show 2D-sketches and simulation snapshots at a temperature of $20\degree$C. Some bases have been added or replaced in the flexible joint and sticky end (depicted in red). Depending on the configuration the average gap size $\langle d_1\rangle$ and $\langle d_2\rangle$ between the sticky ends and flexible joints changes. This can be used to control $\alpha$, whereas $\beta$ is determined by the number of base pairs in the sticky ends. The purple arrows at the center of panels (\textbf{d}) and (\textbf{e}) highlight the sections that have the identical sequence in the hybridized molecules.}
\label{fig:panel5}
\end{figure*}

We studied the topological features of two fully assembled three-armed $S$-$S'$ molecules connected through sticky-end hybridization. 
We introduce two angles to characterize the relative orientation of the two molecules: the bending angle $\alpha$, which measures the collinearity of the two COM3 the connection points where the arms of $S$ and $S'$ hybridize through their sticky ends, and the angle $\beta$, which is defined as the twist angle between the two dihedral planes of the molecules (assuming they are sufficiently planar). This is schematically shown in Fig. \ref{fig:panel5}(a). 

Both $\alpha$ and $\beta$ are determined by the length of the flexible joint $l$ between the dsDNA arm and the sticky overhang. When $l$ is larger than a critical threshold $l_c$, the connection is rather floppy and $\alpha$ and $\beta$ are independent of the sticky ends. When $l$ is smaller than $l_c$ on the other hand, $\alpha$ and $\beta$ will be directly affected by the length and number of bases that are hybridized in the sticky ends. We would expect that when e.g. $l=0$, i.e. when there is no flexible joint, $\alpha = 180\degree$ and $\beta$ is strictly dominated by the number of base pairs in the sticky ends. 

We conducted simulations with $l = 4, 3, 2, 1$ and $0$ inert thymine bases between the sticky end and the dsDNA arm. To our surprise, none of these systems showed a tendency towards the expected $\alpha$ = 180\si{\degree}. 
Our simulations show (Fig.\ref{fig:panel5}(b)) that the nicks between the sticky ends and one backbone of the dsDNA arms constitute an additional degree of freedom. The DNA exists there in two equilibrium states: a closed state, where the system is tightly bonded and behaves as if there was no nick, and an open state, where the nick produces a kink. This changes drastically the relative orientation of the molecules. Furthermore, we found that the average distance between consecutive phosphates where the nicks are located, $<d_{1}>=1.36$ nm and $<d_{2}>=2.35$ nm respectively , are larger than the equilibrium backbone distance in dsDNA, which is $0.6$~nm. 
In a certain sense, reducing this distance between the terminal arm sites and the sticky ends on either side allows to straighten up the structure to the desired angle $\alpha=180\degree$.

Since the stability of DNA depends mainly on two types of interactions, the stacking between adjacent base pairs and the base-pairing between complementary bases, we relate the weakness of these interactions in the nick to the induction of the kink in the backbone. 
It is well known that the stacking interaction, which drives the coplanar alignment of adjacent bases, is sequence dependent ~\cite{Saenger} and may explain why the distances $\langle d_{1}\rangle $ between A-C bases (with large stacking energy) and $\langle d_{2}\rangle $ between T-T bases (with lower stacking energy) differ (c.f. Fig.\ref{fig:panel5} and ~\cite{Sulc:2012} for the parameterisation of the relative strength of stacking interactions).

 To test this hypothesis we replaced the appropriate bases to produce configurations with the largest possible stacking interaction in the nicks (G-C). The selected bases are shown in red in the 2D-sketch of Fig.\ref{fig:panel5}(c). This design change resulted in a reduction of the relevant distances to $\langle d_{1}\rangle =1.14$ nm and $\langle d_{2}\rangle =0.52$ nm, respectively, which confirms that the stacking interaction in the nicked region is indeed important for the control of the relative orientation of the molecules. It becomes also evident that this adaptation alone is not sufficient for straightening up the $S$-$S'$ structure since the distances on either side of the hybridized sticky ends still differ.

In order to see how the sequence in the dsDNA arms beyond the sticky ends affects the alignment of the hybridized Y-DNA molecules, we simulated the same pair of molecules as in Fig.\ref{fig:panel5}(c) but this time using the same sequence in both oligonucleotides $S_3$ and $S'_3$, shown in Fig.\ref{fig:panel5}(d). We measured $\langle d_{1}\rangle =1.0$ nm and $\langle d_{2}\rangle =0.54$ nm for the two average distances between the terminal backbone site on the sticky end and arm, which is a relative although minor improvement with respect to the previous case and obviously insufficient to achieve a straight alignment. We therefore attributed this remaining discrepancy to the different sequences in the sticky ends. 
Simulations where the sticky ends and the complementary sequences on the arms carry exactly the same sequence for three consecutive nucleotides into each entity (c.f. $S_3$ and $S'_3$ in  Fig.\ref{fig:panel5}(e)) confirmed this assumption and we found that both distances are now on average roughly the same at $\langle d_{1}\rangle = 55$nm and $\langle d_{2}\rangle =0.54$ nm, so practically identical. For this configuration, we found that the average bending angle $\langle \alpha \rangle$ between the arm formed by the $S_{2}-S_{3}$ pair and the arm formed by the $S'_{2} - S'_{3}$ pair is $\langle \alpha \rangle = 161 \pm 10\degree$, hence very close to the desire value of $180\degree$.

We also measured the twist angle $\beta$, taking the configuration in Fig.~\ref{fig:panel5}(e) as initial configuration. In a short, straight B-DNA molecule, the twist angle between two base pairs ($i$ and $i+n$) is strongly correlated with the number $n$ of intermediate base pairs as the local equilibrium twist per base pair is $34.285\degree$ for pitch of approximately 10.5 bp.

First we based the calculation on the angle between the vectors linking the end of the free arms (i.e. those that are not hybridized through the sticky ends) -- $\mathbf{v} = \mathbf{a}_{2}-\mathbf{a}_{1}$ for the $S$ molecule and $\mathbf{v}' = \mathbf{a}'_{1}-\mathbf{a}'_{2}$ for the $S'$ molecule. However, this led to rather inconclusive values due to the large thermal fluctuations of the free arms that continuously alter the orientation of the dsDNA arms. We therefore adopted a different procedure to determine the average twist angle.

To minimize fluctuations we kept one of the molecules as static anchor, for example $S'$, and measured the twist angle between two base pairs that are situated 5 bp away from the core of the corresponding molecule in the hybridized part of the arms.
This may appear somewhat \textit{ad hoc}, but note that when $S$ and $S'$ molecules form part a complex network and are hybridized on all and not only two of their arms, the fluctuations would be greatly reduced. The results are summarized in Table~\ref{table:beta}.

\begin{table}[h]
\centering
\begin{tabular}{c c}
\toprule
\hspace{0.5cm} $n$ \hspace{0.5cm} & $\langle \beta \rangle$ \\
\midrule
33           &  $60 \pm	12.4$  \degree       \\
34           &  $81	\pm 13.4$  \degree       \\
35           &  $97 \pm 12.7$  \degree       \\
36           &  $122 \pm 12.9$ \degree       \\
\bottomrule
\end{tabular}
\caption{Average twist angle $\langle\beta\rangle$ between two base-pairs located at a distance of $n$ bp away, along the complementary hybridized arms. The configuration with $n=33$ (0 extra base-pairs added to the middle region in between the two nicks) corresponds to the one of Fig.~\ref{fig:panel5}(e)}.
\label{table:beta}
\end{table}

In the case described above, there are $n=33$ bp between the two selected base pairs, this is $33/10.5 = 3.1428$ helical turns or what it is the same, a local twist angle of $0.1428 \times 360\degree=51.42\degree$. From our simulation, we obtained $\langle\beta \rangle =60 \pm 12.4 \degree$, which is in good agreement with the theoretical value.
After adding extra base pairs to central regions of the sticky ends between the two nicks, we saw also reasonable agreement within the error ranges for 1 and 2 additional bp ($n=34$ and $35$), but not for 3 bp ($n=36$), where we expected a local twist of $154\degree$.

To cross-check these results we ran simulations where both $S$ and $S'$ molecules were integrated dynamically (i.e. $S'$ was not anymore a static anchor) and after equilibration subsequently drained of all kinetic energy by performing Langevin dynamics at a temperature quench to $T=0\degree$K. Under these conditions the correlation of the local twist with the number of intermediate base pairs was in perfect agreement with the theoretical case of about $34.285\degree$ per base pair, leading us to the conclusion that this design principle for the twist angle $\beta$ is basically correct, but difficult to verify on the two-molecule level.

%%%%%%%%%%%%%%%%%%%%%%%%%%%%%%%%%%%%%%
%%%   Section 4: Closing remarks   %%%
%%%%%%%%%%%%%%%%%%%%%%%%%%%%%%%%%%%%%%
\section{\label{sec:conclusion}Closing Remarks}

The overall aim of this paper is to demonstrate that the coarse-grained model of oxDNA can be a helpful guide in the design of a multi-armed DNA structure by providing significant insight into its sequence-dependent topological features. In particular, we studied the three-armed DNA building blocks that were used to construct DNA hydrogel systems as described in Ref~\cite{ZhongyangXing:2018} and applied modifications on these building blocks by adding inert bases in the original design. We first looked at the individual three-armed structure and proposed a measure to quantify the planarity ($d_p$) and arm orientations ($\theta_{1}$, $\theta_{2}$ and $\theta_{3}$) as key parameters to characterize the topological features of the $S$ and $S'$ molecules. By conducting the simulations, we found that the previous assumption the three-armed system are `flat' and the dsDNA arms equally separated is somewhat idealized. Furthermore, we observed a `kink' in the twisted central core, which may give rise to non-canonical base-pairing, an aspect that is beyond the remit of oxDNA and would require more detailed atomistic models. It should be noted though that setting up an atomistic representation of the multi-valent conformations is far from trivial. We propose to add inert bases in the central core region to promote a more defined hybridization between complementary nucleotides. The structural behavior of these modified systems were recorded and further studied. The melting curves measured by simulations are in full accordance with the experimental results, proving that the oxDNA model can be successfully applied to study multi-valent DNA structures that are significantly more complex than B-DNA. 

We also studied the relative geometry of two three-armed DNA structures that are hybridized via complimentary sticky ends. We used two parameters, a bending angle $\alpha$ and a twist angle $\beta$, to define the straightness and relative rotation of the linked components as building blocks of the DNA hydrogel. We observed that even for systems without a flexible joint the bending angle $\alpha$ is still not equal to $180\degree$, contrary to our initial expectation. This turned out to be caused primarily by the flexibility that is introduced through the nicks between the interrupted phosphate backbones. This insight inspired us to introduce DNA ligase that can seal these open nicks in the future if needed. We found that the straightness of the connected region depends also on the symmetry of the sequence of the sticky ends and arms and that palindromic sequences stabilize the arrangement of the two hybridized molecules towards a straighter alignment. The twist angle $\beta$ we expect to be determined by the length of the sticky ends where for instance 10 base pairs should lead to roughly $\beta = 360\degree$. This could be confirmed in runs at temperature $T=0\degree$K where we drained all thermal energy off. At finite temperatures the correlation between the number of base pairs in the sticky ends and the twist angle $\beta$ could be only verified for a couple of additional base pairs. This suggests that due to the large thermal fluctuations and the nicks in the interrupted backbones at the sticky ends the twist angle $\beta$ may be more difficult to measure in this way.  

Although our present study forms only a relatively small portion of the multi-valent systems, our observations can be easily extended to more complex systems and are relevant for the viscoelastic properties of the DNA hydrogel on larger length scales. The topological features we were able to extract allow us to parameterize structural key parameters in even more coarse-grained models more accurately. It forms also a starting point for the development of design principles of DNA-based hydrogels, which we will derive in future work. 

\acknowledgements{
YAGF acknowledges support from Consejo Nacional de Ciencia y Tecnología (CONACyT) PhD Grant No. 384582.
MH was supported by the Research Interns@Strathclyde program. 
OH acknowledges support from the EPSRC Early Career Fellowship Scheme (EP/N019180/2). 
This work used the ARCHER UK National Supercomputing Service (http://www.archer.ac.uk) and the ARCHIE-WeSt High Performance Computer (www.archie-west.ac.uk) based at the University of Strathclyde.}

\bibliography{dna_hydrogel}

\beginappendix

\section*{Appendix}

%%%%%%%%%%%%%%%%%%%%%%%%%%%%%%%%%%%
\subsection{Experimental details}
%%%%%%%%%%%%%%%%%%%%%%%%%%%%%%%%%%%

All the single-stranded DNA used to hybridize the three-armed structures were purchased from Integrated DNA Technologies, Inc. in dry state without further purification. The dry materials were later suspended in deionised water separately and stored in fridge at $4\degree$C for further use.  To measure the melting/hybridizing curves of a three-armed DNA structure, we mixed up equal-molar of the three corresponding single-stranded DNA that can form such structure (e.g. $S_1$, $S_2$ and $S_3$ for $S$-DNA).  NaCl was also added to the sample to facilitate the DNA hybridization by screening the negative charges from DNA backbones. The resulting solution contained the concentration of the three-armed DNA at around $0.8~\mu$M with [NaCl] = $200$~mM.  
The melting curves of $S$-DNA, $S’$-DNA, and its variants were pictured by measuring the $260$ nm peak absorbance of DNA mixture solution using UV-vis spectroscopy equipped with a temperature controlling system.  The temperature varied from $10\degree$C to $75\degree$C for  the measurement shown in Fig.\ref{fig:panel3}(a) and from $20\degree$C to $80\degree$C for that shown in Fig.\ref{fig:panel4}(a) at the rate of $0.2\degree$C/min.  The absorbance curves were then normalized for comparison.

%%%%%%%%%%%%%%%%%%%%%%%%%%%%%%%%%%%
\subsection{Changing the oligonucleotides sequence to stabilize the nicks \label{olinuclsequence}}
%%%%%%%%%%%%%%%%%%%%%%%%%%%%%%%%%%%

As mentioned in section \ref{sec:effectFJ}, the relative orientation between two complementary Y-DNA molecules $T$ and $T'$ depends on their sequence. For example, the presence of the free joint can in turn add a lot of flexibility to the complex in the section where two molecules hybridize (see Fig.~\ref{fig:panel1}(c)). The sequence of these molecules (made by 138 nucleotides each) is given in Table~\ref{table:sequence}.

\begin{figure}[htbp]
	\centering
	\includegraphics[width=0.5\textwidth]{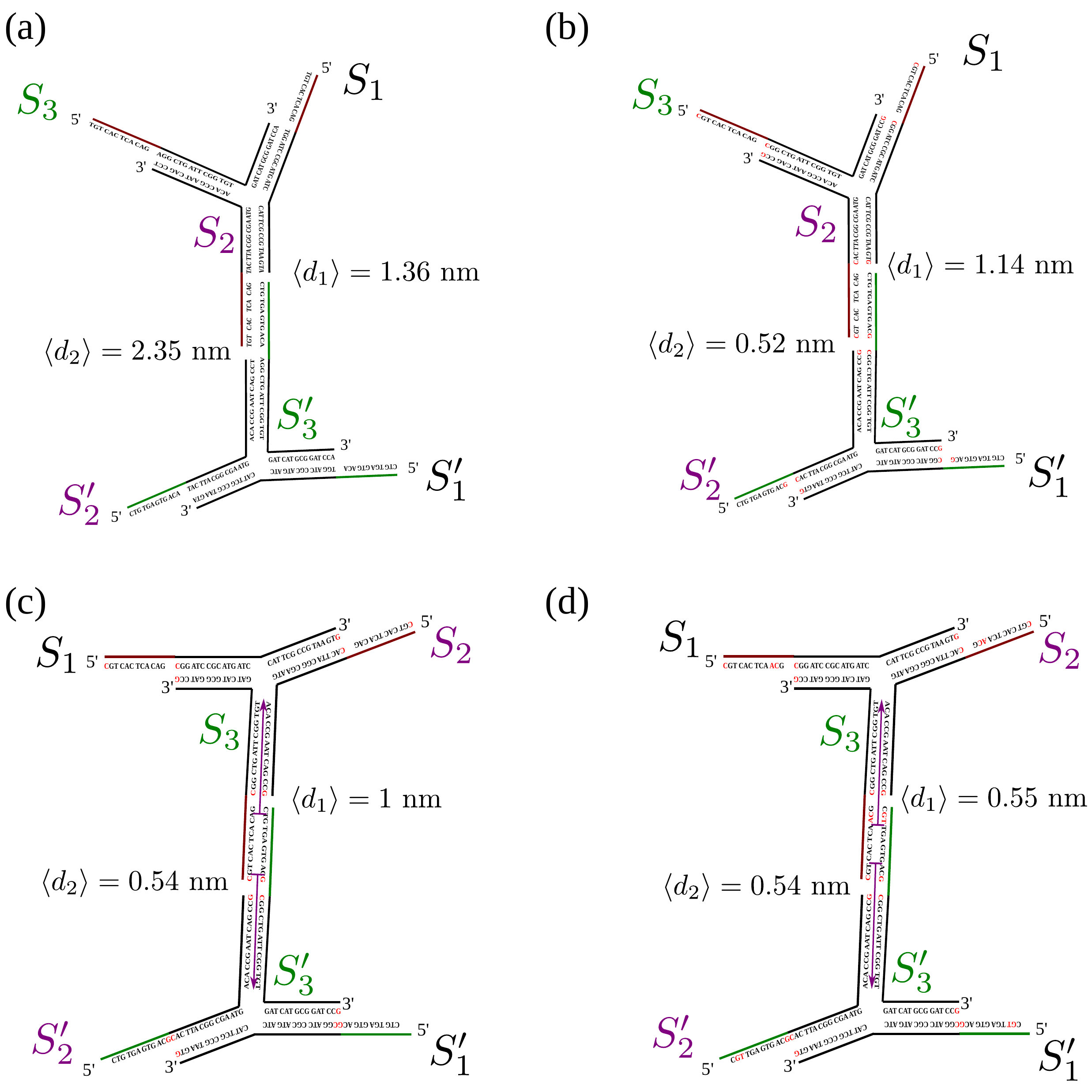}
	\caption{(\textbf{a}) 2D-sketch of the two molecules without flexible joint. Panels (\textbf{b})-(\textbf{d}) show 2D-sketches with the respective sequence in which some bases have been replaced (depicted in red). Depending on the configuration the average gap size $\langle d_1\rangle$ and $\langle d_2\rangle$ in the nicks changes. The purple arrows at the center of panels (\textbf{c}) and (\textbf{d}) highlight the sections that have the identical sequence in the hybridized molecules. The sequences shown in panels (a)-(d) correspond to the 3D snapshots in Fig.\ref{fig:panel5}(b)-(e)}
	\label{fig:S1}
\end{figure}

Remarkably, when the flexible joint was removed, we found in our simulations that the DNA segments that link the two molecules were not aligned. This means that the angle $\alpha$ between the two oligonucleotides that are hybridized by their sticky ends, differs from the expected value of $180\degree$ (expected for such short segments of DNA, 42 bp in total: 15 bp in each of the two hybridized arms and 12 bp in the sticky end). The 3D snapshot from the simulations, for the case in which two complementary molecules are joint through the sticky ends in oligonucleotide $S_{2}$ and $S_{3}^{\prime}$ is shown in Fig.~\ref{fig:panel5}(b). The 2D sketch for the same case is shown in Fig.~\ref{fig:S1}(a). The sequence of this system is the same as the one in Table~\ref{table:sequence} but without the four nucleotides in the free joint, so each molecule is made by 126 nucleotides.

This discrepancy in the angle $\alpha$ was related to the presence of kinks, located exactly where there is a discontinuity (or nick) in the backbone where the two sticky ends meet. In fact, when we measured the average distance between consecutive phosphates in the nicks ($<d_{1}>$=1.36 nm and $<d_{2}>$=2.35 nm respectively), we found that this was larger than the equilibrium backbone distance of dsDNA (roughly 0.6 nm). Therefore, reducing this distance can be seen as a way to align the molecules to form the desire configuration with $\alpha = 180\degree$. A possible way to achieve this, is by increasing the base-pair and stacking interactions between the nucleotides surrounding the nicks, in order to stabilize the DNA duplex formed by the hybridization of bases in this region. Since both of these interactions are larger for GC nucleotides, we replaced part of the original sequence by G's and C's. This is shown schematically in Fig.~\ref{fig:S1}(b) and Table~\ref{table:seqb}, where the replaced nucleotides are shown in red.

\begin{table}[h!]
\centering
\resizebox{\columnwidth}{!}
{
\begin{tabular}{*4c}
\toprule
ssDNA    & Sticky end      & \multicolumn{2}{c}{Core} \\
\midrule
 {}      &        {}       &        Segment I   &         Segment II   \\
 %PART OF THE TABLE FOR THE T-MOLECULE
 \begin{tabular}{@{\ }l@{}}
 $S_{1}$ \\ $S_{2}$ \\ $S_{3}$
 \end{tabular}

 & $5^{\prime}-$ \textcolor{red}{C}GTCACTCACAG &  

 $\left\{\begin{tabular}{@{\ }l@{}}
 \textcolor{red}{C}GGATCCGCATGATC \\ \textcolor{red}{C}ACTTACGGCGAATG \\ \textcolor{red}{C}GGCTGATTCGGTGT  \end{tabular}\right.$ & 

 \begin{tabular}{@{\ }l@{}}
    CATTCGCCGTAAGT\textcolor{red}{G} $-3^{\prime}$ \\ ACACCGAATCAGCC\textcolor{red}{G} $-3^{\prime}$ \\ GATCATGCGGATCC\textcolor{red}{G} $-3^{\prime}$  \end{tabular}\\
         &                 &                     &                       \\
 %PART OF THE TABLE FOR THE T'-MOLECULE
 \begin{tabular}{@{\ }l@{}}
 $S_{1}^{\prime}$ \\ $S_{2}^{\prime}$ \\ $S_{3}^{\prime}$
 \end{tabular}

 & $5^{\prime}-$ CTGTGAGTGAC\textcolor{red}{G} & 

 $\left\{\begin{tabular}{@{\ }l@{}}
 \textcolor{red}{C}GGATCCGCATGATC \\ \textcolor{red}{C}ACTTACGGCGAATG \\ \textcolor{red}{C}GGCTGATTCGGTGT  \end{tabular}\right.$ & 

 \begin{tabular}{@{\ }l@{}}
    CATTCGCCGTAAGT\textcolor{red}{G} $-3^{\prime}$ \\ ACACCGAATCAGCC\textcolor{red}{G} $-3^{\prime}$ \\ GATCATGCGGATCC\textcolor{red}{G} $-3^{\prime}$  \end{tabular}\\

\bottomrule
\end{tabular}
}
\caption{The sequence provided in this table corresponds to the 2D sketches of Fig.\ref{fig:S1}(b)-(c) and the 3D snapshots of Fig.~\ref{fig:panel5}(c)-(d). Since the magnitude of the stacking and base-pair interactions is larger for GC nucleotides, we used them to replace the corresponding nucleotides at the nicks. With this, we found that the average distance between consecutive phosphates where the nicks are located decreased from $\langle d_{1} \rangle = 1.36$ nm and $\langle d_{2} \rangle = 2.35$ nm to $\langle d_{1} \rangle = 1.14$ nm and $\langle d_{2} \rangle = 0.52$ nm, for the case where the molecules are attached by the $S_{2}$ and $S_{3}^{\prime}$ oligonucleotides.}
\label{table:seqb}
\end{table}

With this modifications the distance at the nicks decreased to $\langle d_{1}\rangle =1.14$ nm and $\langle d_{2}\rangle =0.52$ nm respectively. However, this was not enough to produce the desire conformation, so we explored how the sequence in the dsDNA arms beyond the sticky ends affects the geometry of the system. To this end we joined the molecules in Table~\ref{table:seqb} through the sticky ends in oligonucleotide $S_{3}$ and $S_{3}^{\prime}$ (instead of $S_{2}$ and $S_{3}^{\prime}$). In this scenario, the base-pair sequence from the nicks to the core of the molecules (indicated by the purple arrows in the 2D sketch of Fig.~\ref{fig:S1}(c)) is exactly the same. This time we found $\langle d_{1} \rangle = 1$ nm and $\langle d_{2}\rangle = 0.54$ nm, a minor improvement (towards finding the configuration with $\langle d_{1} \rangle$ = $\langle d_{2} \rangle$) with respect to the previous case. 

The difference between the values of $\langle d_{1} \rangle$ and $\langle d_{2} \rangle$ could only be related to the difference in the sticky ends sequence. However, the question remained on how many base pairs had to be replaced in this section to equalize the distances in the nicks. For this we modified the base sequence of the sticky ends in the vicinity of the nicks to match that of the sticky end close to the nick on the opposite backbone, one base pair at a time until we found the sequence shown in Fig.~\ref{fig:S1}(d) and Table~\ref{table:seqd}. The outcome was that the nucleotide sequence of the three base-pairs closer to the nicks (and in the direction of the sticky ends) is important to obtain the same average distances $\langle d_{1} \rangle = 0.55$ nm and $\langle d_{2} \rangle = 0.54$ nm.

\begin{table}[h!]
\centering
\resizebox{\columnwidth}{!}
{
\begin{tabular}{*4c}
\toprule
ssDNA    & Sticky end      & \multicolumn{2}{c}{Core} \\
\midrule
 {}      &        {}       &        Segment I   &         Segment II   \\
 %PART OF THE TABLE FOR THE T-MOLECULE
 \begin{tabular}{@{\ }l@{}}
 $S_{1}$ \\ $S_{2}$ \\ $S_{3}$
 \end{tabular}

 & $5^{\prime}-$ \textcolor{red}{C}GTCACTCA\textcolor{red}{AC}G &  

 $\left\{\begin{tabular}{@{\ }l@{}}
 \textcolor{red}{C}GGATCCGCATGATC \\ \textcolor{red}{C}ACTTACGGCGAATG \\ \textcolor{red}{C}GGCTGATTCGGTGT  \end{tabular}\right.$ & 

 \begin{tabular}{@{\ }l@{}}
    CATTCGCCGTAAGT\textcolor{red}{G} $-3^{\prime}$ \\ ACACCGAATCAGCC\textcolor{red}{G} $-3^{\prime}$ \\ GATCATGCGGATCC\textcolor{red}{G} $-3^{\prime}$  \end{tabular}\\
         &                 &                     &                       \\
 %PART OF THE TABLE FOR THE T'-MOLECULE
 \begin{tabular}{@{\ }l@{}}
 $S_{1}^{\prime}$ \\ $S_{2}^{\prime}$ \\ $S_{3}^{\prime}$
 \end{tabular}

 & $5^{\prime}-$ C\textcolor{red}{GT}TGAGTGAC\textcolor{red}{G} & 

 $\left\{\begin{tabular}{@{\ }l@{}}
 \textcolor{red}{C}GGATCCGCATGATC \\ \textcolor{red}{C}ACTTACGGCGAATG \\ \textcolor{red}{C}GGCTGATTCGGTGT  \end{tabular}\right.$ & 

 \begin{tabular}{@{\ }l@{}}
    CATTCGCCGTAAGT\textcolor{red}{G} $-3^{\prime}$ \\ ACACCGAATCAGCC\textcolor{red}{G} $-3^{\prime}$ \\ GATCATGCGGATCC\textcolor{red}{G} $-3^{\prime}$  \end{tabular}\\

\bottomrule
\end{tabular}
}
\caption{The sequence provided in this table ensures that the distance between nucleotides at the nicks is the same ( $\langle d_{1} \rangle = \langle d_{2} \rangle$). It corresponds to the 2D sketch of Fig.~\ref{fig:S1}(d) and the 3D snapshot of Fig.~\ref{fig:panel5}(e).}
\label{table:seqd}
\end{table}

%%%%%%%%%%%%%%%%%%%%%%%%%%%%%%%%%%%%%
\subsection{Planarity of the molecules}
%%%%%%%%%%%%%%%%%%%%%%%%%%%%%%%%%%%%%

In order to compute the variable $d_{p}$ related to the planarity of the $T$ and $T'$ molecules, we need first to define the direction of each of the dsDNA arms of a nanostar molecule. This is done in the following way:

\begin{itemize}
\item Starting from the core of the molecule, we identify the center of the first non-denatured base pair along each arm.
\item Then we compute the COM of these three points; i.e. find COM3.
\item The unitary vectors ($\mathbf{a}_{i}$ with $i=1,2,3$) pointing from COM3 to the end of each arm, define the direction of the arms.
\end{itemize}

We can find two vectors ($\mathbf{A}$ and $\mathbf{B}$) lying on the plane that passes through the end of the three arms, by subtracting two of the previously defined vectors. For example:

\begin{subequations}
\begin{align}
    \mathbf{A} &= \mathbf{a}_{2} - \mathbf{a}_{1}. \\
    \mathbf{B} &= \mathbf{a}_{3} - \mathbf{a}_{1}.
\end{align}
\end{subequations}

The cross product $\mathbf{n} = \mathbf{A} \times \mathbf{B}$ defines the vector $\mathbf{n}$, normal to the plane. The general equation of this plane is:

\begin{equation}
n_{x}x + n_{y}y + n_{z}z = d,
\label{eq:genplane}
\end{equation}

\noindent where $d$ is a constant and $n_{x}$, $n_{y}$ and $n_{z}$ represent the Cartesian components of $\mathbf{n}$. Since all the points ($x, y, z$) on the plane satisfy Eq.~\ref{eq:genplane}, this includes the components of any of the $\mathbf{a}_{i}$ vectors. Therefore, with the components of for example $\mathbf{a}_{2}$, i.e. ($a_{2x}$, $a_{2y}$, $a_{2z}$), the value of the constant $d$ can be found:

\begin{equation}
d = a_{2x} \; n_{x} + a_{2y} \; n_{y} + a_{2z} \; n_{z}.
\end{equation}

Finally, the distance $d_{p}$ from the plane to COM3 (with coordinates ($c_{x}$, $c_{y}$, $c_{z}$)) can be computed with the formula:
\begin{equation}
d_{p} = \frac{n_{x} c_{x} + n_{y} c_{y} + n_{z} c_{z} + d}{\sqrt{n^{2}_{x} + n^{2}_{y} + n^{2}_{z} }}.
\end{equation}

\end{document}